\documentclass[12pt,preprint]{aastex}

\slugcomment{{\copyright} 2006, The American Astoronomical Society.}
\begin{document}

\title{
  Distribution of dust clouds around the central engine of NGC 1068%
  \thanks{
    Based on data collected at Subaru Telescope, which is
    operated by the National Astronomical Observatory of Japan.
  }
}

\author{Daigo Tomono and Hiroshi Terada}
\affil{Subaru Telescope, National Astronomical Observatory of Japan\\
  650 North A`ohoku Place, Hilo, Hawaii 96720, U.S.A.}
\email{tomono at subaru.naoj.org}
\and
\author{Naoto Kobayashi}
\affil{Institute of Astronomy, Graduate School of Science, University of Tokyo\\
  2-21-1 Osawa, Mitaka, Tokyo 181-0015, Japan}

\begin{abstract}
  We studied the distribution of dust clouds around the central
  engine of NGC 1068 based on shifted-and-added 8.8 --
  12.3~{\micron} (MIR) multi-filter images and
  3.0 -- 3.9{~\micron} ({\it L}-band) spectra obtained with the Subaru
  Telescope.
  In a region of 100~pc (1{\farcs}4) around the central peak,
  we successfully constructed maps of color temperatures and
  emissivities of the MIR and {\it L}-band continua as well as
  the 9.7~{\micron} and 3.4~{\micron} dust features
  with spatial resolutions of 26~pc (0{\farcs}37) in the MIR and
  22~pc (0{\farcs}3) in the {\it L}-band.
  Our main
  results are: 1) color temperature of the MIR continuum scatters around
  the thermal equilibrium temperature with the central engine as the heat
  source while that of the {\it L}-band continuum is higher and
  independent upon distance from the central engine;
  2) the peak of the 9.7~{\micron} silicate absorption feature is
  shifted to a longer wavelength at some locations;
  3) the ratio of the optical depths of the dust features
  is different from the Galactic values and show complicated spatial
  distribution; and
  4) there is a pie shaped warm dust cloud as an enhancement in the
  emissivity of the MIR continuum extending about 50~pc to the north
  from the central engine.
  We speculate that
  material falls into the central engine through this cloud.
\end{abstract}

\keywords{
      galaxies: active
  --- galaxies: individual (\objectname[M 77]{NGC 1068})
  --- galaxies: nuclei
  --- galaxies: Seyfert
  --- infrared: galaxies
}

\section{INTRODUCTION}
\label{sec:Introduction}

Physics around black holes, including accretion of material and launching of
jets, has been a fundamental question in astronomy for a long time
\citep[e.g.][]{Eardley1975}.
It is not yet clearly known how accreting material disposes of its
angular momentum while falling into a black hole, and how and where a jet
is launched from around
a black hole.
With a jet of 100~pc scale observed \citep[e.g.][]{Gallimore1996},
active galactic nuclei (AGNs) of Seyfert galaxies are the best targets to
attack the mystery 
of accreting material and jets.

There are two classes of Seyfert galaxies; Seyfert 1 and Seyfert 2.
They have similarities and differences in, for example, their visible spectra,
which are commonly known to be explained with
the unified model \citep{Antonucci1993}.
According to the unified model, an AGN has a black hole at its heart.
Material falls down into the black hole through an accretion disk while
emitting X-ray and UV light.
Throughout this paper, we
define the system of the black hole and the
accretion disk as
``central engine'' following \cite{Galliano2003}.
The unified model proposes
three classes of clouds
around the central engine.
First, narrow line regions (NLRs) spread over $\sim$ 100~pc and emit
permitted and forbidden lines with velocity widths of $\la 1000$~km~sec$^{-1}$.
Second, broad line regions (BLRs) in the vicinity of the central
engine emit permitted lines of $\sim 10000$~km~sec$^{-1}$ velocity widths.
Third, an opaque compact but geometrically thick dusty torus around the central
engine hides the direct view to the central engine and their BLRs from our line
of sight when the central engine is viewed through the torus.
The dusty torus is the key to the unified model because it is
the one which distinguishes Seyfert 1 and Seyfert 2:
Seyfert 2 nuclei are viewed through the dusty
torus and BLRs are only seen in scattered light.

NGC 1068 is one of the nearest Seyfert 2 galaxies.
Many observations have revealed various features that confirm the
unified model;
an accretion disk of 1~pc scale \citep{Gallimore1997,Gallimore2001},
the Seyfert 1 nucleus seen in polarized light \citep{Antonucci1985},
and NLRs \citep{Evans1991}.
However, no clear image of the dusty torus has been obtained
\citep[e.g.][]{Bock2000,Tomono2001,Rouan2004}.
Only recently, \cite{Jaffe2004} found a 2.1~pc scale emission component
of warm (320~K) dust surrounding a more compact hot ($> 800$~K) dust
component with the Mid-Infrared Interferometric Instrument coupled to
the Very Large Telescope Interferometer.
They identified the hot component as the inner wall of the dusty torus.
It is a step towards clear images of the dusty torus including its
temperature structure, density distribution, and relations with
structures in larger and smaller scales.

On the other hand, the inner radii of the dusty tori are being measured with
reverberation mapping.
This method measures lag time of light curves of an AGN in the {\it
V}-band and that in the {\it K}-band.
In reverberation mapping, the measured lag time is interpreted as the time
duration in which light from the AGN reaches the inner wall of the dusty torus.
Multiplying the lag with the speed of light, the inner radius of the dusty
torus $R_{in}$ can be obtained.
\cite{Minezaki2004} found a good correlation between lags and absolute
magnitudes in the {\it V}-band in a number of Seyfert 1 galaxies.
It implies that $R_{in}$ is governed by luminosity of the central engine.
Adopting their results,
minimum $R_{in}$ for NGC 1068 is estimated to be
$\sim$~0.045~pc from $m_V \sim$~11.5~mag \citep{Sandage1973}.
The estimated radius is consistent with the upper limit of 1~pc
measured by \cite{Jaffe2004}.
The method seems to be promising in unveiling the innermost structures of dusty
clouds around AGNs.
However, we have to note that the results do not necessarily require the
dusty clouds around the AGNs to be in the shape of a torus.
It is still critical to obtain direct images of dusty tori.

It is best to look for a dusty torus in the infrared (IR).
First,
from thermal equilibrium, it is expected that dust in a torus of 10~pc
scale radiates mainly around 10~{\micron} \citep{Pier1992}.
Next,
there are absorption/emission features of carbonaceous dust at
3.4~{\micron} and of silicate dust at 9.7~{\micron}.
Comparison of
the dust features
may be used to probe
the temperature structure of a dusty
torus as has been suggested by \cite{Imanishi2000}.
Because {\it L}-band emission is dominated by dust at $\sim
1000$~K, the 3.4~{\micron} absorption feature should
originate from the dust clouds
in a few pc from the central
engine \citep{Pier1992}.
On the other hand, the 9.7~{\micron} absorption feature only measures
the extinction towards the outer region of the dusty torus because dust
of $\sim 300$~K,
which is the dominant source of the mid-infrared (MIR) emission, is located
further out from the heating source.
Therefore, the combination of MIR spectra (8~--~13~{\micron})
and {\it L}-band spectra (2.8 -- 4.2~{\micron}) would
probe the temperature distribution of dust in the line of sight through
the dusty torus.

This paper reports the results from observations of
NGC 1068 in the MIR and in the {\it L}-band with the
Subaru Telescope, which is best suited for this study with the superb
image quality in IR.
Two instruments, the Mid-Infrared Test
Observation System (MIRTOS) and the Infrared Camera and Spectrograph
(IRCS), were used with the fine pixel scales optimized for the image
quality of the telescope.
The remainder of this paper is organized as follows: Sections
\ref{sec:MIR} and \ref{sec:NIR} describe observations and data reduction
in the MIR and in the {\it L}-band, respectively;
Section \ref{sec:Discussion} discusses emission mechanisms of the continua,
spectral and spatial change of the dust features, and distributions of
warm dust clouds; and {\S} \ref{sec:Conclusion} summarizes the results.
Throughout this paper, we assume the redshift of $0.0038$
for NGC 1068 \citep{Bottineli1990}. One arcsecond corresponds to 72~pc at
the distance of NGC 1068, assuming the Hubble
constant of $H_0 = 75$~km~sec$^{-1}$~Mpc$^{-1}$.
\cite{Galliano2003} measured positions of the various spatial structures
in the NGC 1068 nucleus
observed at different wavelengths and found that the central engine is
at the IR peak.
We employ their registration to compare the results in the MIR and in the
{\it L}-band as well as those in the literature.

\section{MIR DATA}
\label{sec:MIR}

\subsection{Observations}

MIR imaging observations of NGC 1068 with the ``silicate filters''
($\lambda/\Delta\lambda \sim 10$) at 7.7, 8.8, 9.7, 10.4, 11.7, and
12.3~{\micron} were made on 1999 December 31, 2000 January 9, and 2000
January 18 UT.
We used the MIRTOS \citep{Tomono2000,Tomonoetal2000} mounted at the
Cassegrain focus of the 8.2-m Subaru Telescope.
Pixel scale was 0{\farcs}067, corresponding to a projected
distance on the sky of 4.8~pc at the distance of NGC 1068.
Sequences of 96 images of 31~msec integration were obtained.
Between the sequences, the telescope was nodded typically 30{\arcmin} to
the north to acquire background images.
In this study, the data described in \cite{Tomono2001} were reduced again
with the higher computing power available today.

\subsection{Shift-and-add and registration of images}
\label{sec:MIRreduction}

Many short-exposure images with 31~msec integration time were
shifted-and-added to construct high-resolution images in each filter.
First, background images, which were made with averaging the
short-exposure images in the nodding pair, were subtracted from the raw
images.
Second, bad pixels and detector array specific features were corrected.
Third, the peak position of the object in each image was measured with a
precision of one tenth of a pixel with Gaussian fitting.
The FWHM of the Gaussian was fixed to be the same as that of the Airy
pattern to pick up the brightest speckle.
Fourth, each pixel of the images was divided to 10 times 10 pixels
resulting with a pixel scale of 0{\farcs}0067.
Fifth, each image of the smaller pixel scale was shifted by registering
the peak position measured in the third step.
Finally, the shifted images were averaged after some images with low
peak values were removed.
Table \ref{tab:minteg} summarizes total integration time of the
short-exposure images used and the ratio of the images used to those
acquired.

\begin{deluxetable}{rrr}
\tablecolumns{3}
\tablewidth{0pc}
\tablecaption{Total integration time and usage rate of short-exposure images
  in MIR}
\label{tab:minteg}
\tablehead{
  \colhead{Wavelength} & \colhead{Total integration time of images used} & \colhead{Percentage of images used} \\
  \colhead{{\micron}} & \colhead{sec} & \colhead{\%}
}
\startdata
  7.7 &  2.7 & 23 \\
  8.8 &  5.2 & 43 \\
  9.7 &  4.9 & 41 \\
 10.4 & 24.9 & 42 \\
 11.7 & 53.7 & 53 \\
 12.3 & 68.2 & 47 \\
\enddata
\end{deluxetable}

The flux was calibrated with the standard star $\alpha$~Ari
\citep{Cohen1999}.
Flux uncertainty was estimated at about 9\% from the measurement of the
other standard star $\alpha$~CMa \citep{Cohen1995}.

\begin{figure}
  \plotone{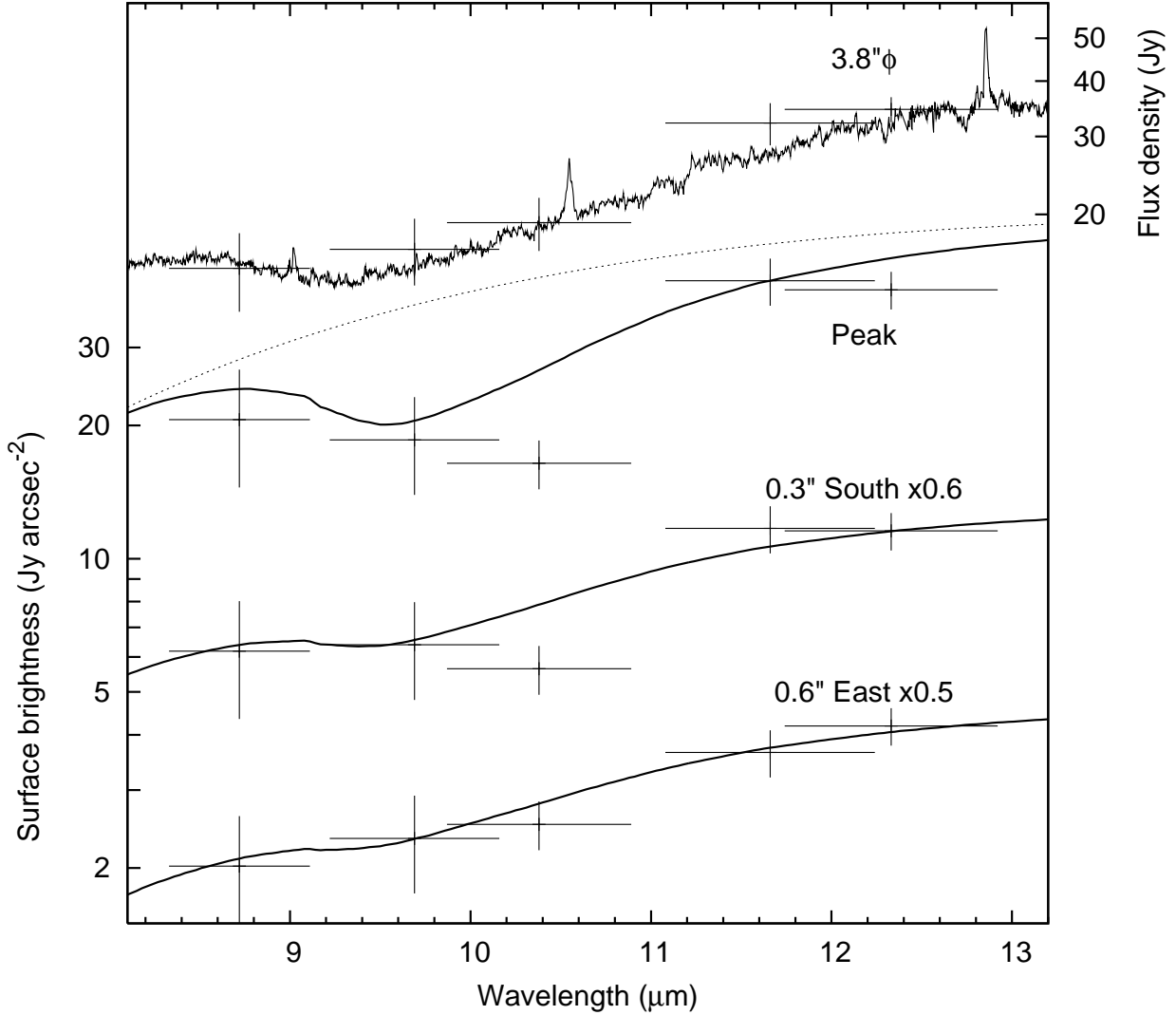}
  \caption[SED and results of grey body fitting in the MIR]
  {SED and results of grey body fitting in the MIR (lower data points,
    curves and left axis)
    and flux densities in a 3{\farcs}8 aperture (upper data points
    and right axis) compared with the
    spectrum measured with the {\it ISO}-SWS \citep[right axis]{Sturm2000}.
    Each spatial element is measured in an area 0{\farcs}15 long
    (P.A. 3{\arcdeg})
    by 0{\farcs}3 wide corresponding to the {\it L}-band slitlet.
    Each data point is shown with the filter path band as a horizontal
    error bar and flux uncertainty of 1$\sigma$,
    including uncertainty in the flux conversion factor,
    as a vertical error bar.
    The thick curves display the resulting grey body models (eq.
    [\ref{equ:MIRmodel}]) on respective spatial elements.
    The dotted curve shows the underlying model continuum for the peak.
    The 10.4~{\micron} data are not used in fitting
    (see {\S} \ref{sec:10.4}).
  \label{fig:mirspec}}
\end{figure}

The shifted-and-added image of each filter was registered and convolved
to have the same FWHM of $\sim$ 0{\farcs}37 (26~pc) as follows.
The peak position of the object for each filter was measured by Gaussian
fitting of the pixels with flux more than 75\% of the peak.
The images were registered assuming that the peaks for each
filters are at the same location on the sky
\citep{Galliano2003}.
The images are then Gaussian convolved so that all the images
have a similar FWHM.
Figure \ref{fig:mirspec} shows the spatially integrated SED
within a 3{\farcs}8 diameter aperture on the registered
narrow-band images.
The {\it ISO}-SWS spectrum in the 14{\arcsec} $\times$ 27{\arcsec}
aperture \citep{Lutz2000,Sturm2000} is also shown for comparison.
The flux density measured from the MIRTOS images are consistent with the
flux density of the continuum in the SWS spectrum.

\subsection{Grey body fitting}
\label{sec:MIRfitting}

We fitted the MIR spectral energy distribution (SED)
with a grey body emission model including the
silicate feature to estimate the temperature and amount of MIR emitting dust,
as well as optical depth of the silicate absorption/emission feature.
It should be noted that the extracted parameters do not represent the
entire physical conditions but only characterize distributions of dust
in some aspects as described below.
The observed SED was fitted with a model flux
\begin{equation}
  F_{\lambda}(\lambda) =
    \varepsilon_{cont}
    \left( \frac{\lambda}{10{\mbox{~\micron}}}\right)^{-n}
    B_{\lambda}(T_{cont},\lambda) \times
    \exp[ -\tau_{9.7} \times k(\lambda) ],
\label{equ:MIRmodel}
\end{equation}
where $B_{\lambda}(T,\lambda)$ is the Planck function of
temperature $T$ and $k(\lambda)$ is the optical depth of the silicate
feature obtained from the IR excess towards $\mu$~Cep
\citep{Roche1984} and normalized at 9.7~{\micron}.

The $\mu$~Cep extinction curve is used in this work to
compare the silicate feature with the 3.4~{\micron} feature as has been
done by \cite{Roche1984}.
The 7.7~{\micron} image was not used because the IR excess data does not
cover the short wavelength.
The peak wavelength of the silicate absorption feature seen in the SEDs
in Figure \ref{fig:mirspec} is different from that observed towards
$\mu$~Cep: the peak wavelength of the absorption is shifted from
9.7~{\micron} to 10.4~{\micron} especially at the central peak.
Because of this, the 10.4~{\micron} image was not used in the SED
fitting.
See {\S} \ref{sec:10.4} for a discussion on the change of the peak
wavelength.

\begin{figure}
  \plotone{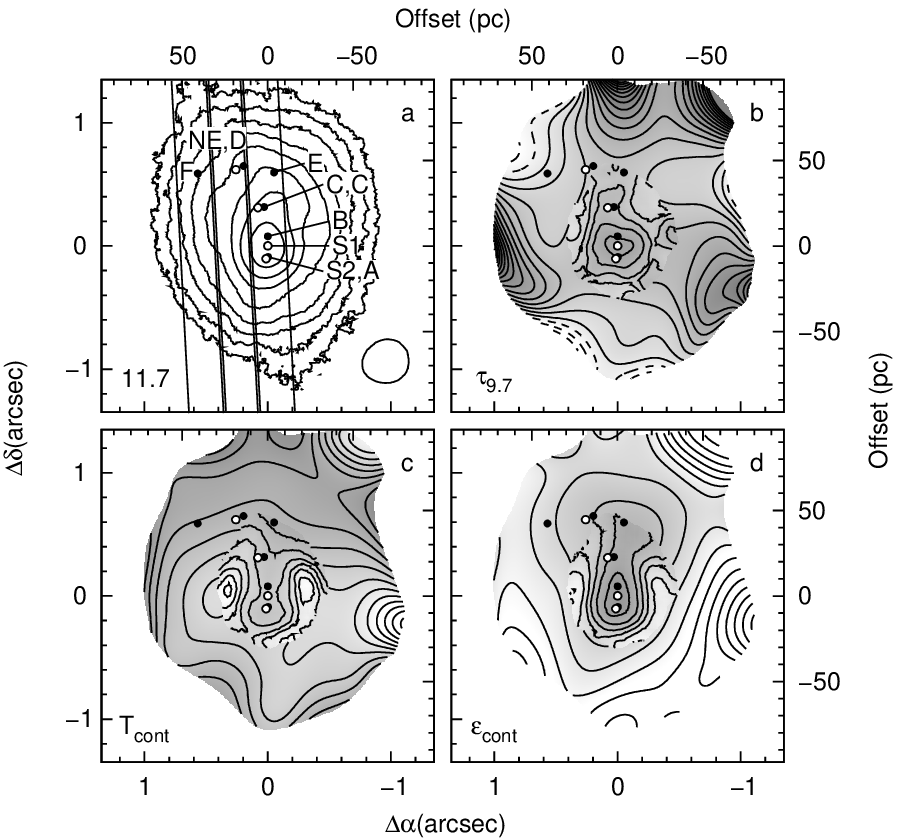}
  \caption[MIR image and grey body parameters]
  {MIR image and grey body parameters (eq. [\ref{equ:MIRmodel}]).
    The fit for the outer region is performed on images convolved with a
    Gaussian of FWHM same as the original PSF (the circle in panel {\it a}).
    Locations of the 5~GHz sources \cite[open circles]{Gallimore1996}
    and the [\ion{O}{3}] clouds \cite[filled circles]{Evans1991} are
    indicated.
    ({\it a}) 11.7~{\micron} image shown with 5 contours in dex.
    The straight lines show
    positions and widths of the three {\it L}-band slits.
    ({\it b}) $\tau_{9.7}$ map.
    Solid contours show $\tau_{9.7}$ between 0.1 and 1.5 (absorption;
    darker grey scale) with intervals of 0.1 while dashed contours show
    $\tau_{9.7}$ of 0 and $-$0.1 (light).
    ({\it c}) $T_{cont}(\mbox{MIR})$ shown with contours between 160~K
    (dark) and 280~K (light) with intervals of 10~K.
    ({\it d}) $\varepsilon_{cont}$ shown with five contours in each dex
    between 10$^{-1.4}$ (dark) and 10$^{-3.8}$ (light).
  \label{fig:MIRTOS}}
\end{figure}

The variables $T_{cont}$, $\varepsilon_{cont}$, and $\tau_{9.7}$ were
treated as free parameters for the fitting.
Color temperature $T_{cont}$ represents the temperature of warm
dust grains emitting the MIR continuum (see {\S} \ref{sec:MIRcont} for a
discussion).
The factor $\varepsilon_{cont}$ is a
product of dust emissivity at 10~{\micron} and beam filling
factor of warm dust grains of $T_{cont}$.
Throughout this paper, we call $\varepsilon_{cont}$ as ``emissivity''.
The mass of the warm dust grains per unit area can be estimated
as $M_{dust}= \varepsilon_{cont} / K_{abs}$, where $K_{abs}$ is
the absorption cross section per unit mass of dust grains.
The dust model \citep{Weingartner2001,Draine2003a} with $R_V = 5.5$
tabulates the absorption cross section
at 10~{\micron} as $K_{abs} = 0.40 \mbox{~cm}^2\mbox{~g}^{-1}$.
Although the dust cloud might be clumpy,
clumpiness does not affect the mass estimation
when $\varepsilon_{cont} < 1$ and a homogeneous temperature
is assumed of warm dust that is emitting continuum.
The other free parameter, $\tau_{9.7}$, is the optical depth of the silicate
feature at 9.7~{\micron} towards the warm dust that is emitting continuum.
A negative optical depth in our model represents emission of the silicate
feature.
Although the spectral index of dust emissivity $n$ is usually
uncertain ranging from 1 to 2, we assumed $n=1.6$ following
\cite{Cameron1993}.
Figure
\ref{fig:MIRTOS} shows spatial distributions of the parameters
fitted on each pixel.
The images were successfully fitted over a region of 50~pc
to the north and 40~pc to the south
from the central peak:
uncertainty in fitted parameters are $\Delta\tau_{9.7} < 1$,
$\Delta T_{cont} < T_{cont}/2$, and
$\Delta\varepsilon_{cont} < \varepsilon_{cont}/2$.
Using the secondary reference star $\alpha$~CMa rather than
$\alpha$~Ari, resulting T$_{cont}$ is within 1.5~K, $\tau_{9.7}$ becomes
about 0.1 deeper, and $\varepsilon_{cont}$ becomes 10\% stronger.
To improve the signal-to-noise ratio of the parameters in the area
surrounding the central peak, the data were also fitted after being
convolved to have 1.4 times wider FWHM.

\cite{Maiolino2001} pointed out that ground based
measurements of the silicate absorption feature at 9.7~{\micron} might
suffer from contamination in continuum with the
unidentified infrared band (UIB) emission at 7.7~{\micron}.
However, \cite{Sturm2000} detected little emission of 7.7~{\micron} UIB
towards NGC 1068 with {\it ISO}-SWS in the 14{\arcsec} $\times$
27{\arcsec} aperture.
The 3.3~{\micron} UIB feature is not seen in the spectrum obtained
towards NGC 1068 by \cite{Imanishi1997} with the 3{\farcs}8 aperture
nor in the {\it L}-band spectra in this work.
In fact, the UIBs detected with ISOCAM \citep{LeFloch2001} originate
almost exclusively from the starburst regions, peaking at a distance of
1~kpc (14{\arcsec}) from the central engine.
They indicated that the AGN contributes less than $\sim$5\% to the total
integrated UIB emission.
Therefore, we concluded that the contamination by the UIBs to the
continuum is negligible in our field of view of of $\sim$200~pc
(2{\farcs}7) around the central engine of NGC 1068.

High temperature areas seen 0{\farcs}34 from the central engine on
east and west sides
coincide with the second and third diffraction rings
(0{\farcs}29 and 0{\farcs}41 radius) of the Airy pattern at 8.8~{\micron}.
Between these diffraction rings,
there are also diffraction rings at longer wavelengths
which might affect the fitting.
Nevertheless, a fitting on convolved Airy images does not reproduce a
peak in fitted temperature.
Therefore, we concluded that the high temperature region is not an
artifact induced by the Airy rings.

Flux was also spatially integrated over the area of each {\it L}-band slitlet
(see {\S} \ref{sec:L-reduction})
and
the integrated flux is fitted with the same model as described in eq.
[\ref{equ:MIRmodel}].
With comparable spatial resolutions achieved for the MIR shifted-and-added
images and
the {\it L}-band seeing limited spectra, no further convolution or
deconvolution was performed.
Results are shown in Figure \ref{fig:mirspec}.

\section{{\it L}-BAND DATA}
\label{sec:NIR}

\subsection{Observation}

A seeing limited spectroscopic observation of NGC 1068 in the {\it
L}-band was conducted on 2000 September 24 UT.
We used the IRCS
\citep{Tokunaga1998,Kobayashi2000} mounted at the Cassegrain focus of
the Subaru Telescope.
The {\it L}-band grism for 2.8~--~4.2~{\micron} was used
with the 0{\farcs}3 slit, yielding the spectral resolution
$\lambda/\Delta\lambda \sim$ 400.
The length of the slit was 18{\arcsec}.
Pixel scale was 0{\farcs}058 pixel$^{-1}$.
NGC 1068 was observed with the slit position angle (P.A.) of 3{\arcdeg}
from the north,
with slit positioned at three positions to map the
eastern area of the central peak:
pointing offsets of $\pm$6{\arcsec} along the slit and
0{\farcs}28 and 0{\farcs}56 to the east were applied.
Positions of the slits are illustrated in Figure \ref{fig:MIRTOS}{\it a}.
We call the slit at the central peak as ``C'', and the slits
in the east as ``0.28E'' and ``0.56E'' according to the
offsets to the east in arcseconds.
Eight exposures of 2 seconds on-source integration time were obtained for
each slit position.
The sky condition was photometric throughout the observation.
Airmass to the target was $1.09 \pm 0.01$.
From the size of the image along the slit,
spatial resolution was estimated to be 0{\farcs}31
or 22~pc in FWHM.
As a telluric standard, a spectrum of BS 813 was obtained
at the similar airmass (1.08).
We selected a F-type star as a telluric standard because it is
known that no strong stellar absorption lines are present in
L-band for F-type stars.
For flat fielding, spectra of dawn sky were obtained at the end of the
night.

\subsection{Data Reduction}
\label{sec:L-reduction}

The {\it L}-band data were reduced following the standard procedure:
the 2-dimensional spectrograms were
background subtracted by the dithering pairs,
flat fielded, and corrected for bad pixels using IRAF%
\footnote{
  IRAF is distributed by the National Optical Astronomy Observatories,
  which are operated by the Association of Universities for Research in
  Astronomy, Inc., under cooperative agreement with the National Science
  Foundation.
}.
Subsequently, positional shifts of the spectrograms of each slit
position were measured with
cross correlations and confirmed to be not more than 0.2 pixels
(0$\farcs$01 or 3.2~{\AA}) in both the spatial and dispersion directions.
The spectrograms were shifted in the dispersion direction on the
2-dimensional spectrograms to cancel
wavelength shifts between the slit positions.
Flux of the spectrograms were then calibrated with the
spectrum of the standard star
BS 813, which was assumed to have an effective
temperature of 7244~K \citep{Johnson1983}.
Wavelengths of the spectra were calibrated using cross-correlation
with a model atmospheric transmittance \citep{Rothman1998}.
Then, the spectra in slitlets of 0{\farcs}15 along the slit were extracted to
examine spatial variations.
Finally, the spectra were convolved with a Gaussian with the FWHM of
8.2{\AA} in the dispersion direction to match the spectral resolution to
that of the instrument and improve the signal-to-noise ratio.

\begin{figure}
  \plotone{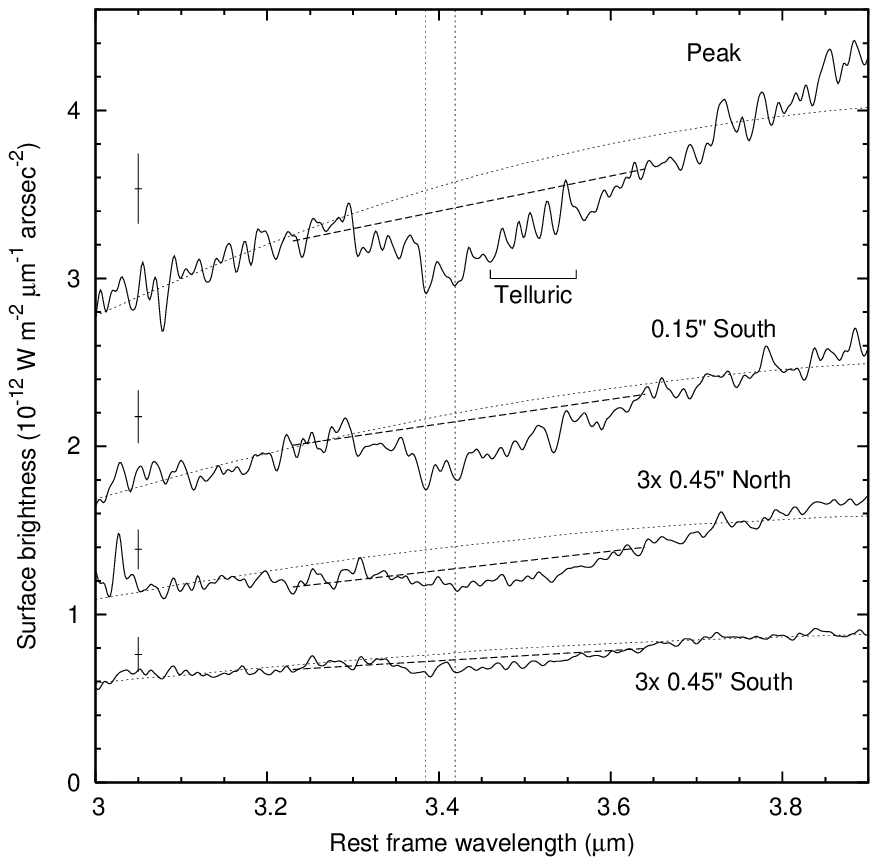}
  \caption[Spectra in the {\it L}-band]
  {{\it L}-band spectra
    on slitlets of 0{\farcs}15 long and 0{\farcs}3 wide.
    Two vertical dotted lines at 3.23~{\micron} (2955~cm$^{-1}$) and
    3.64~{\micron} (2925~cm$^{-1}$) show the location of the sub-peaks of
    the 3.4~{\micron} absorption features.
    The dashed lines connecting continuum levels at 3.23 and
    3.64~{\micron} show continua for measuring $\tau_{3.4}$
    \citep{Sandford1991}.
    The dotted curves show model continuum (eq.
    [\ref{equ:NIRmodel}]) with dust emissivity index of $n = 1.6$,
    fitted for the data between 3.0 and
    3.9~{\micron} excluding the wavelength range of the dust feature
    (3.23 and 3.64~{\micron}).
    The crosses in the left show typical 1$\sigma$ statistical
    uncertainties (vertical length) of each data point and the spectral
    resolution (horizontal length).
    Periodic features between 3.46 and 3.56~{\micron} seen in
    some spectra are artifacts due to telluric absorption
    \citep{Goto2003AO}.
  \label{fig:ngc.separate}}
\end{figure}

Figure \ref{fig:ngc.separate} shows spectra of some slitlets in the
slit C.
It should be noted that the slopes of the continuum at $\lambda <$
3.0~{\micron} and $\lambda >$ 3.9~{\micron} are sensitive to the
atmospheric conditions during the observation.
In the following analysis, we did not use the data in these wavelength
ranges to avoid the uncertainty of the continuum slope.
The periodic feature at 3.46~--~3.56~{\micron} in the spectrum of the
central peak in Figure \ref{fig:ngc.separate} is from the periodic
telluric absorption in this wavelength range.
Because the instrumental line profile changes upon spatial distribution
of the illumination within the slit, the difference of the instrumental
line profiles of the target and the standard star causes the residual
features when the two spectra are divided each other
\citep[see the detailed discussion in][]{Goto2003AO}.

\subsection{The 3.4~{\micron} absorption feature and the {\it L}-band continuum}
\label{sec:3.4}

\begin{figure}
  \plotone{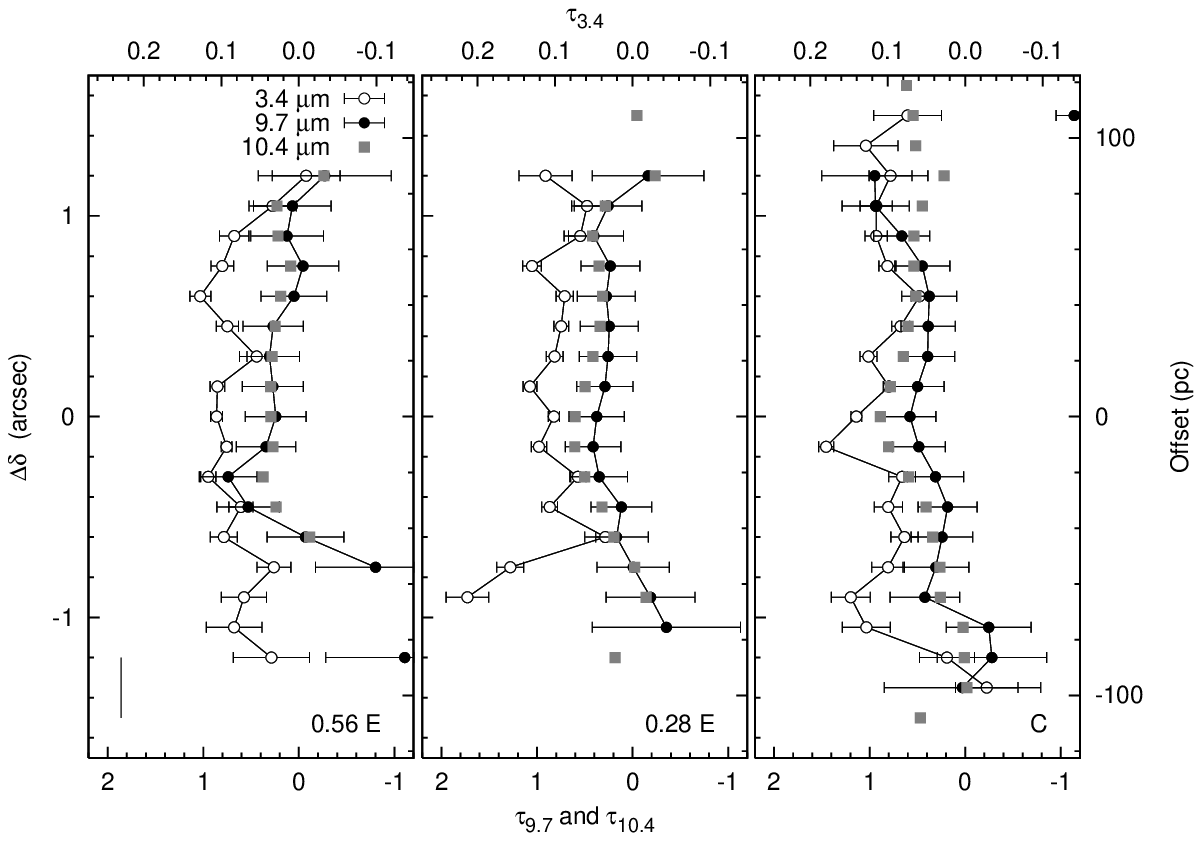}
  \caption[Spatial variation of optical depths of the dust features]
  {Spatial variation of optical depths of the dust features for three
    slit positions.
    Open circles, filled circles, grey filled squares show
    $\tau_{3.4}$, $\tau_{9.7}$, and $\tau_{10.4}$, respectively.
    Positive values of $\tau$ mean absorption while negative values
    mean emission.
    Error bars are the 1$\sigma$ uncertainty.
    Error bars for $\tau_{10.4}$, which are similar to
    that for $\tau_{9.7}$, are not shown.
    The vertical line at bottom left shows the spatial resolution (FWHM) of
    the data.
  \label{fig:tau}}
\end{figure}

The absorption feature around 3.4~{\micron} by carbonaceous dust is
clearly seen in the spectra in Figure \ref{fig:ngc.separate}.
Following \cite{Sandford1991}, optical depth of the dust feature
$\tau_{3.4}$ was measured as follows:
1) the continuum level at the absorption feature was estimated by
connecting the continuum levels at rest wavelengths 3.23 and
3.64~{\micron} with straight line and
2) $\tau_{3.4}$ was derived by averaging the optical depths at two sub
peaks of the dust feature at rest wavelengths 3.38~{\micron}
(2955~cm$^{-1}$) and 3.42~{\micron} (2925~cm$^{-1}$).
The 3.48~{\micron} (2870 cm$^{-1}$) sub peak was not included in the
measurements
because of the residual telluric feature described in {\S}
\ref{sec:L-reduction}.
Figure \ref{fig:tau} shows spatial change of $\tau_{3.4}$ along with
$\tau_{9.7}$ and $\tau_{10.4}$.

\begin{figure}
  \plotone{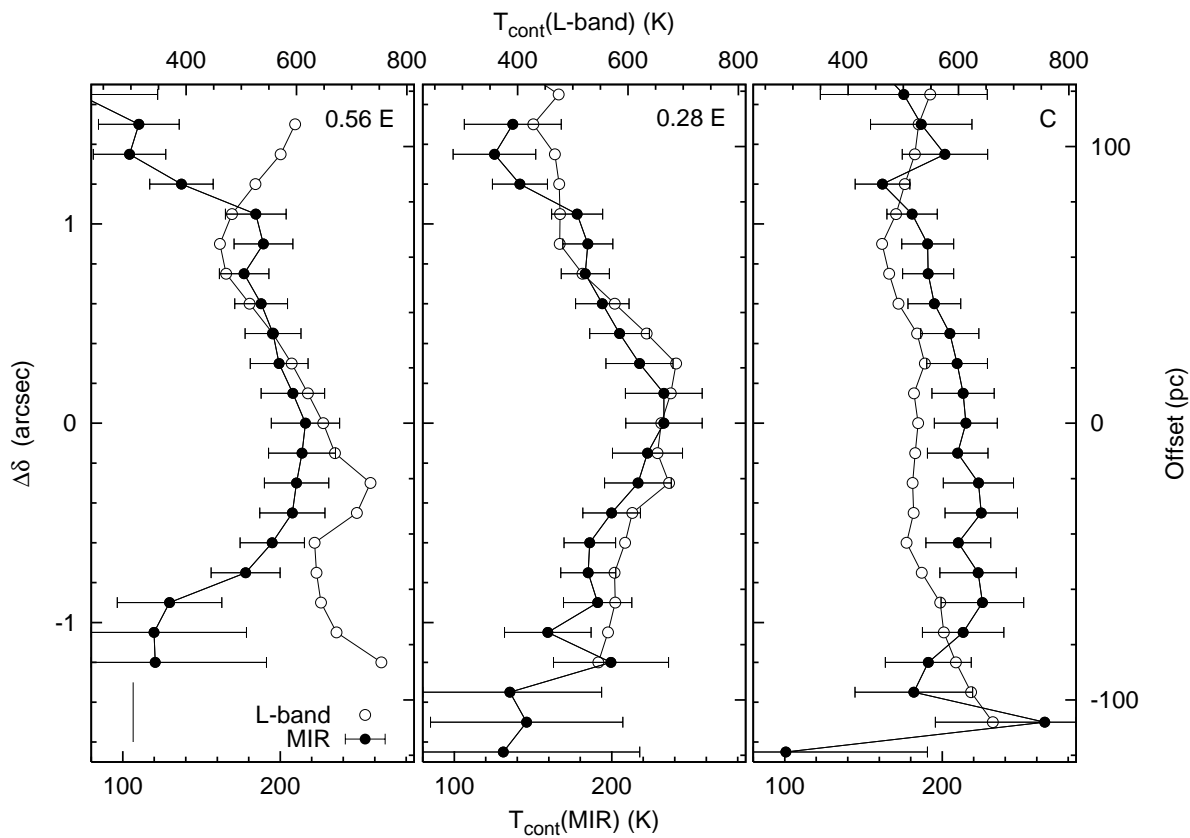}
  \caption[Color temperatures of continua]
  {Color temperatures of the continua in the {\it L}-band (open circles
    and upper abscissa) and in the MIR (filled circles and lower
    abscissa) on the three slit positions.
    Error bars show the 1$\sigma$ uncertainty of each measurement, which are
    less than the size of the symbols for the {\it L}-band measurements.
    The vertical line at bottom left shows the spatial resolution (FWHM) of
    the data.
  \label{fig:Tback}}
\end{figure}

Color temperature $T_{cont}$ and emissivity
$\varepsilon_{cont}$ in the {\it L}-band were estimated by fitting a grey
body model
\begin{equation}
  F_{\lambda}(\lambda) =
    \varepsilon_{cont}
    \left( \frac{\lambda}{10{\mbox{~\micron}}}\right)^{-n}
    B_{\lambda}(T_{cont}, \lambda)
\label{equ:NIRmodel}
\end{equation}
to the spectra at 3.00~--~3.23~{\micron} and 3.64~--~3.90~{\micron}.
These wavelength ranges were chosen to avoid the 3.4~{\micron} dust
feature and the
unstable continuum level at the both ends of the wavelength
range due to changes of the atmospheric conditions.
We assumed a dust emissivity index of $n = 1.6$ for comparison of the
parameters with those in the MIR ({\S} \ref{sec:MIRfitting}).
Spatial change of the color temperature in the {\it L}-band along with
those in MIR are shown in Figures \ref{fig:Tback}.

\section{RESULTS AND DISCUSSION}
\label{sec:Discussion}

\subsection{Emission mechanisms of MIR and {\it L}-band continua}
\label{sec:background_source}

Continua are detected both in the MIR and in the {\it L}-band
over a region of 100~pc to the north and to the south
from the central engine.
Because
\cite{Gratadour2003} detected no evidence of stellar activity in the
vicinity of the central engine in their {\it K}-band spectra,
stellar activity should be negligible as a source of the extended
IR continua.
In the following, we will show that the MIR
continuum is emitted from dust in thermal equilibrium with the UV
radiation from the central engine while 
the continuum in the {\it L}-band is emitted by Very Small Grains (VSGs).

\subsubsection{MIR continuum emission}
\label{sec:MIRcont}

\begin{figure}
  \plotone{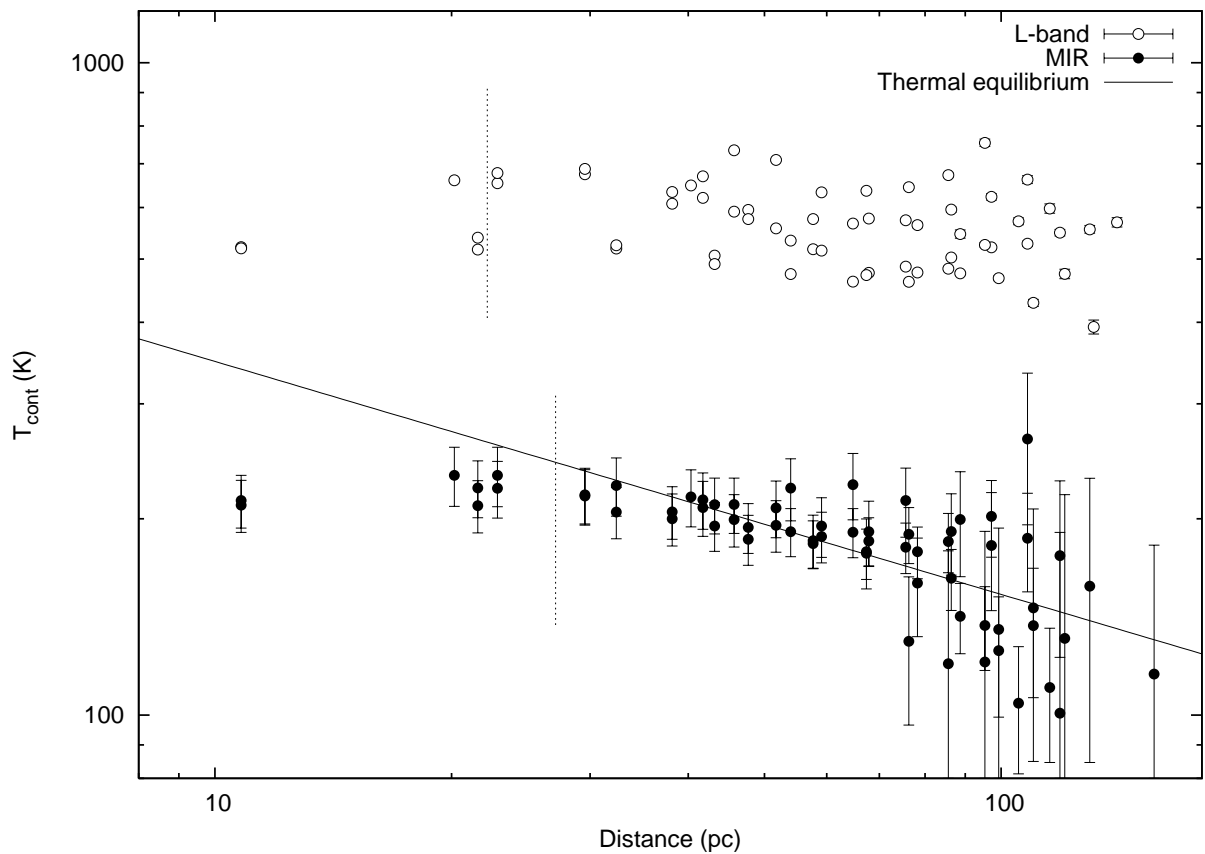}
  \caption[Color temperatures of continua for different distance from
  the central peak]
  {Color temperatures of continua measured on each slitlet in
    MIR (filled circles) and in the {\it L}-band (empty circles).
    Horizontal axis shows projected distance on the sky from the central engine.
    Error bars show the 1$\sigma$ uncertainty of each measurement, almost
    of which are less than the size of the symbols for the {\it L}-band
    measurements.
    The solid line shows thermal equilibrium temperature of dust
    (eq. [\ref{equ:Cameron1993}])
    heated by a UV source of
    $L = 2.2\times10^{11}\mbox{~L$_\sun$}$ \citep{Telesco1980}.
    The spatial resolutions of the data and contamination of light from
    the peak limit the color temperatures to be accurate only to the
    right of the vertical dotted lines (see text).
  \label{fig:Tback.prof}}
\end{figure}

Assuming a dust emissivity index $n$ 1.6,
\cite{Cameron1993} estimated dust temperature $T$ in thermal equilibrium with
a UV radiation field as
\begin{equation}
  R = 0.14 \left(\frac{L}{1.5\times10^{11}\mbox{~L$_\sun$}}\right)^{0.5}
           \left(\frac{T}{1500\mbox{~K}}\right)^{-2.8} \mbox{pc},
\label{equ:Cameron1993}
\end{equation}
where $L$ is the luminosity of the UV source and $R$ is the linear distance
from the heating source.
Figure \ref{fig:Tback.prof} shows color temperature $T_{cont}$ measured
in the MIR and the estimated temperature for $L =
2.2\times10^{11}\mbox{~L$_\sun$}$
\citep[at luminosity distance of 15~Mpc]{Telesco1980}
as a function of the distance from the central engine.
The distance is assumed to be the same as the projected
distance on the sky.
The vertical dotted lines in the Figure show spatial resolutions of data
in the MIR and in the {\it L}-band, which are different because of
different wavelengths and observation conditions.
Because the light from the central engine, of which color temperature
might be different from that for the extended continua, might
contaminate the continua, the color temperatures measured
around the central engine within the spatial resolutions are not
accurate.

In Figure \ref{fig:Tback.prof},
the measured $T_{cont}$ in the MIR scatters
around the estimated equilibrium
temperature.
This implies that, as a whole, the MIR emitting dust
is mainly heated by the
central engine and is in thermal equilibrium
as has been suggested by \cite{Galliano2005}.

\subsubsection{{\it L}-band continuum emission}
\label{sec:NIRcont}

We interpret that the extended {\it L}-band continuum is emitted by VSGs
among the following three possibilities.
First, the continuum is not dominated by the scattered light of the
continuum from the central engine.
\cite{Packham1997} measured spatially resolved polarization degrees in the
{\it J}, {\it H}, and {\it K}-bands around the central engine and found
that they are not more than than 5\%.
Second,
the {\it L}-band continuum is not emitted from dust in thermal
equilibrium.
As shown in Figure \ref{fig:Tback.prof}, color temperature of the {\it
L}-band continuum is higher than expected for dust in thermal
equilibrium.
Third, color temperature in the {\it L}-band
remains roughly constant as far as 100~pc from the central
engine;
suggesting that hot grains emitting {\it L}-band continuum
exist in this area even if their column density is not as high as those
emitting MIR continuum.
As \cite{Rouan2004} and \cite{Gratadour2005} suggested,
the uniformity of the color temperature can be explained as emission from
VSGs,
which emit continuum with color temperature higher than their
equilibrium temperature due to
thermal fluctuation
upon absorption of UV photons \citep{Sellgren1984,Sellgren1985}.

\begin{figure}
  \plotone{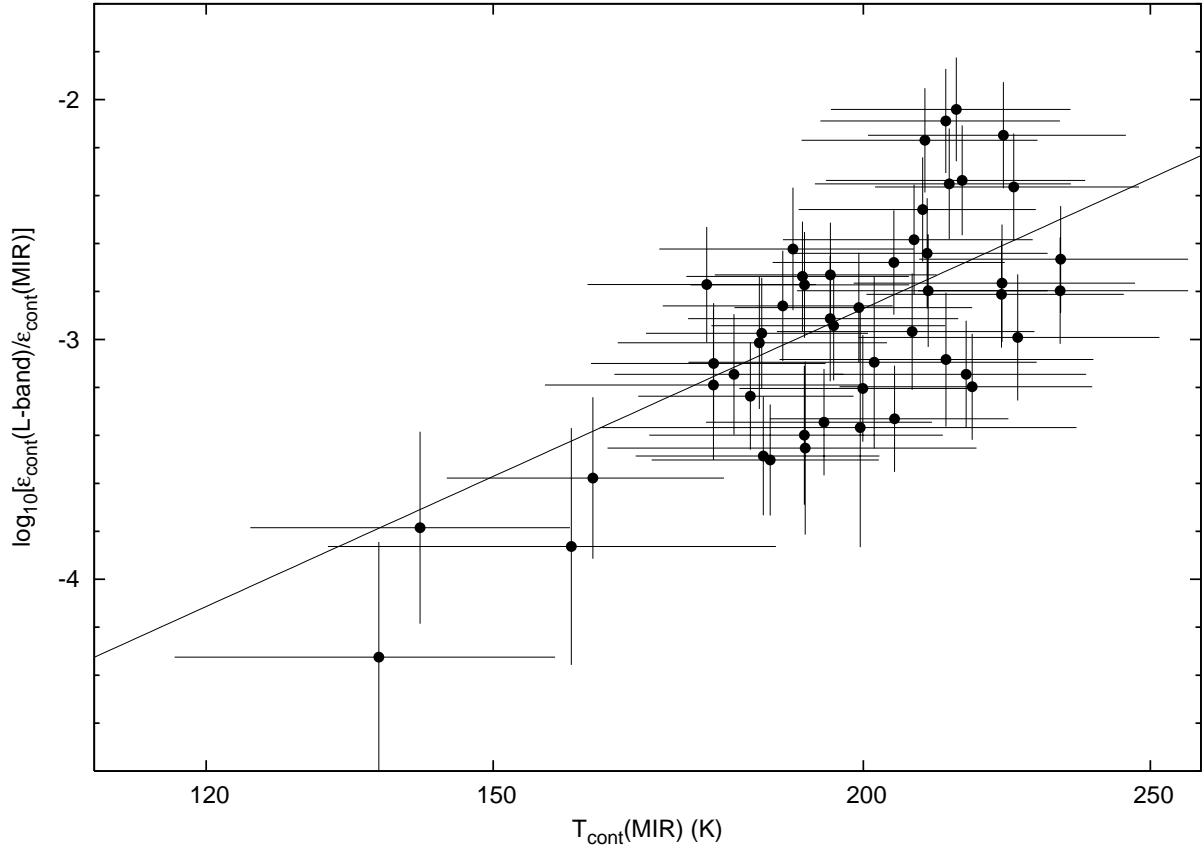}
  \caption[Ratio of emissivities]
  {Ratio of emissivities
    measured in the MIR and in the {\it L}-band
    compared with color temperature measured in the MIR.
    Error bars are the 1$\sigma$ uncertainty of each measurement.
    The solid line shows the relation
    $\varepsilon_{cont}(\mbox{{\it L}-band}) / \varepsilon_{cont}(\mbox{MIR})
    = ( T_{cont}(\mbox{MIR})/ 651\mbox{~K} )^{5.6}$
    (eq. [\ref{equ:eratio}]).
  \label{fig:eratio}}
\end{figure}

If the continuum is emitted by VSGs,
the incident UV photon flux does not determine color temperature but
determines brightness, or emissivity in our grey body model, of the continuum.
We actually found the
change of emissivity of the {\it L}-band continuum upon incident UV
photon flux
from our grey body fittings, suggesting that the {\it L}-band
continuum is indeed emitted by VSGs.
Figure \ref{fig:eratio}
shows correlation between
$\varepsilon_{cont}(\mbox{{\it L}-band})/\varepsilon_{cont}(\mbox{MIR})$
and $T_{cont}(\mbox{MIR})$.
Thermal equilibrium temperature
$T_{cont}(\mbox{MIR})$ should correlate with the UV photon flux.
On the other hand,
since $\varepsilon_{cont}$ is proportional to column density of dust
grains emitting the continuum,
$\varepsilon_{cont}(\mbox{{\it L}-band})/\varepsilon_{cont}(\mbox{MIR})$
gives the ratio of dust grains excited to be VSGs to
those in thermal equilibrium emitting continuum in the MIR.
In more detail,
energy $F$ of UV photons absorbed by dust grains is balanced with the
grey body emission with temperature $T$ as follows:
\begin{equation}
  F = \pi \int
    \varepsilon_{cont} \left( \frac{\lambda}{10{\mbox{~\micron}}}\right)^{-n}
    B_\lambda( T, \lambda ) \; d\lambda
  \propto \varepsilon_{cont} T^{4+n},
\label{equ:T4+n}
\end{equation}
where $\varepsilon_{cont}$ is the emissivity of the dust cloud at
10~{\micron}.
Division of two equations for the {\it L}-band and MIR continua
leads to the following equation:
\begin{equation}
  \frac{\varepsilon_{cont}(\mbox{{\it L}-band})}{\varepsilon_{cont}(\mbox{MIR})}
  = \frac{F(\mbox{{\it L}-band})}{F(\mbox{MIR})} \times
    \left[ \frac{T_{cont}(\mbox{MIR})}{T_{cont}(\mbox{{\it L}-band})} \right]
    ^{4+n}.
\label{equ:eratio}
\end{equation}
The solid line in Figure \ref{fig:eratio} shows the fitting result when
assuming $n = 1.6$ and adjusting the factor
  $T_{cont}(\mbox{{\it L}-band}) \times
  [ {F(\mbox{MIR})/F(\mbox{{\it L}-band})}
  ] ^{1/(4+n)}$
as a free parameter.
Relation between the measured $\varepsilon_{cont}(\mbox{{\it
L}-band})/\varepsilon_{cont}(\mbox{MIR})$ and the measured
$T_{cont}(\mbox{MIR})$ in Figure \ref{fig:eratio} roughly matches with the
assumption in the range of almost two orders of
$\varepsilon_{cont}(\mbox{{\it L}-band})/\varepsilon_{cont}(\mbox{MIR})$.
The fitted factor (651~K) and the observed $T_{cont}(\mbox{{\it L}-band}) \sim
560$~K leads to $F(\mbox{{\it L}-band}) / F(\mbox{MIR}) \sim 0.43$,
suggesting that VSGs absorb almost half of the UV energy that is absorbed by
dust grains in thermal equilibrium.
The uniformity of the ratio $F(\mbox{{\it L}-band}) / F(\mbox{MIR})$
suggests that dust in thermal equilibrium and VSGs are mixed well around
the central engine.

\subsection{Spectral profile change of the silicate absorption feature}
\label{sec:10.4}

At 10.4~{\micron}, the silicate absorption towards the central engine is
deeper than expected from the standard silicate feature.
Figure \ref{fig:mirspec} compares the observed SEDs with expected
spectra from the IR excess towards $\mu$~Cep \citep{Roche1984}.
There are a number of similar wavelength shift of the silicate feature
in the literature.
The interferometric spectrum by \cite{Jaffe2004} also shows a shift of
the peak of the absorption feature towards longer wavelength.
Silicate emission feature detected towards a type 1 AGN \citep{Sturm2005}
and quasars \citep{Siebenmorgen2005} is shifted to a longer wavelength.

The spatially integrated SED in Figure \ref{fig:mirspec} is consistent
with the spectrum observed through the 14{\arcsec}$\times$20{\arcsec}
aperture of the {\it ISO}-SWS \citep{Sturm2000} which shows the normal
spectral profile of the silicate absorption feature.
This suggests that the spectral profile is normal as a whole and there
is a spatial change of the profile.

To measure the optical depth of the feature without being affected by
the change, we fitted the MIR SED with a gray body emission model
without using the 10.4~{\micron} image (\S \ref{sec:MIRfitting}).
If the 10.4~{\micron} image were included in the grey body fitting, the
resulting $\tau_{9.7}$ would have been higher than expected from the
literature \citep{Roche1984b, Tomono2001}, especially near the central
engine where $\tau_{9.7}$ would have to be significantly adjusted for
the low flux at 10.4~{\micron}.

\begin{figure}
  \plotone{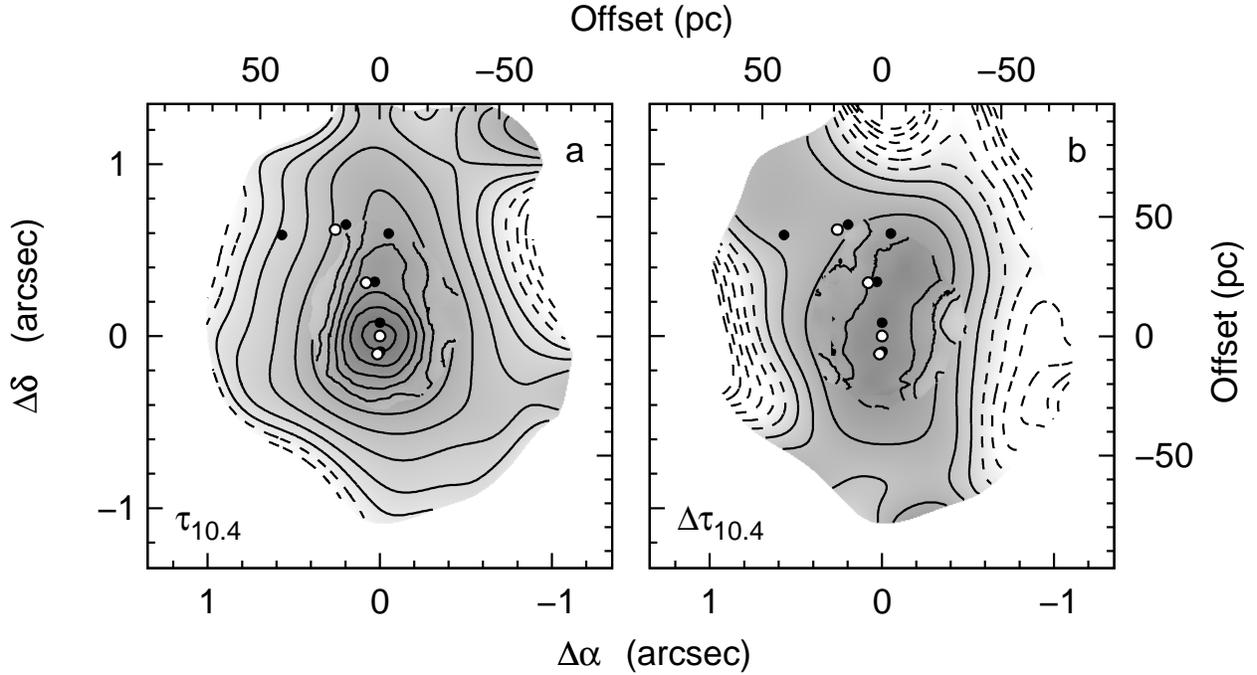}
  \caption[Spatial distribution of the silicate absorption feature
    at 10.4~{\micron} and 9.7~{\micron}]
  {Spatial distribution of the silicate absorption feature at
    10.4~{\micron} and 9.7~{\micron}.
    ({\it a}) Optical depth measured at 10.4~{\micron}.
    Solid contours show $\tau_{10.4}$ between 0.1 and 0.9 (absorption;
    stronger with darker grey scale) with intervals of 0.1 while dashed
    contours show $\tau_{10.4}$ between 0 and $-$0.3 (emission; light).
    ({\it b}) Excess optical depth at 10.4~{\micron}
    ($\Delta\tau_{10.4}$; see text) for the $\mu$~Cep extinction curve.
    Solid contours show $\Delta\tau_{10.4}$ between 0.05 and 0.35
    (larger $\tau_{10.4}$; darker grey scale) with intervals of 0.1
    while dashed
    contours show $\Delta\tau_{10.4}$ below $-$0.05 with intervals of 0.1.
    Locations of the 5~GHz sources \cite[open
    circles]{Gallimore1996} and the [\ion{O}{3}] clouds \cite[filled
    circles]{Evans1991} are indicated.
  \label{fig:silicates}}
\end{figure}

Figure \ref{fig:silicates}{\it a} shows spatial distribution of
optical depth in the 10.4~{\micron} filter $\tau_{10.4}$.
It is measured with comparing the 10.4~{\micron} image with the model
continuum surface brightness fitted from the other images.
The shape of the 10.4~{\micron} absorption area is different from the
9.7~{\micron} absorption area (Figure \ref{fig:MIRTOS}{\it b}).
To illustrate the difference, we compared the measured $\tau_{10.4}$
with that expected from the measured $\tau_{9.7}$ and the IR excess
towards $\mu$~Cep \citep{Roche1984}.
Figure \ref{fig:silicates}{\it b} shows the excess optical depth at
10.4~{\micron} ($\Delta\tau_{10.4}$) near the central engine and to the
north north east.

The change of the spectral profile might be due to change of dust
properties as suggested by \cite{Siebenmorgen2005} and \cite{Sturm2005}
for type 1 AGNs and quasars.
The area with large $\Delta\tau_{10.4}$ in Figure \ref{fig:silicates}{\it b}
coincides with the pie shaped high
emissivity area ({\S} \ref{sec:pie}) seen in Figure
\ref{fig:MIRTOS}{\it d}.
This suggests that the spectral profile change of the silicate
absorption feature occurs in dense regions.

\subsection{Spatial variation of 9.7~{\micron} and 3.4~{\micron}
absorption features}
\label{sec:dust}

It is known that optical depth of the absorption feature of silicate
dust and that of carbonaceous dust are proportional towards Galactic
sources.
\cite{Roche1984} showed that $A_V/\tau_{9.7} = 18.5$ towards Wolf-Rayet
stars and B supergiants in our Galaxy.
\cite{Pendleton1994} measured optical depths of the carbonaceous dust
and concluded that $A_V/\tau_{3.4}$ is $\sim 150$ for sources in the
Galactic center or $\sim 250$ for sources in the local interstellar
matter (ISM).
From these results, ratio of the optical depths of the absorption
features $\tau_{9.7}/\tau_{3.4}$ is $\sim$~8.1 for the sources in the
Galactic center or $\sim$~13.5 for the local ISM.
Towards AGNs, $\tau_{9.7}/\tau_{3.4}$ is expected to be smaller because
of temperature gradient in the dusty torus in the line of sight
\citep{Imanishi2000}.
In fact, they measured $\tau_{3.4}$ towards NGC 1068 in the 3{\farcs}8
aperture, compared it with $\tau_{9.7}$ measured in the 4{\farcs}7
aperture by \cite{Roche1984b}, and yielded $\tau_{9.7}/\tau_{3.4} = 4.3$.
With the data with higher spatial resolution, spatial variation of the
ratio might show a trace of the dusty torus.

\begin{figure}
  \plotone{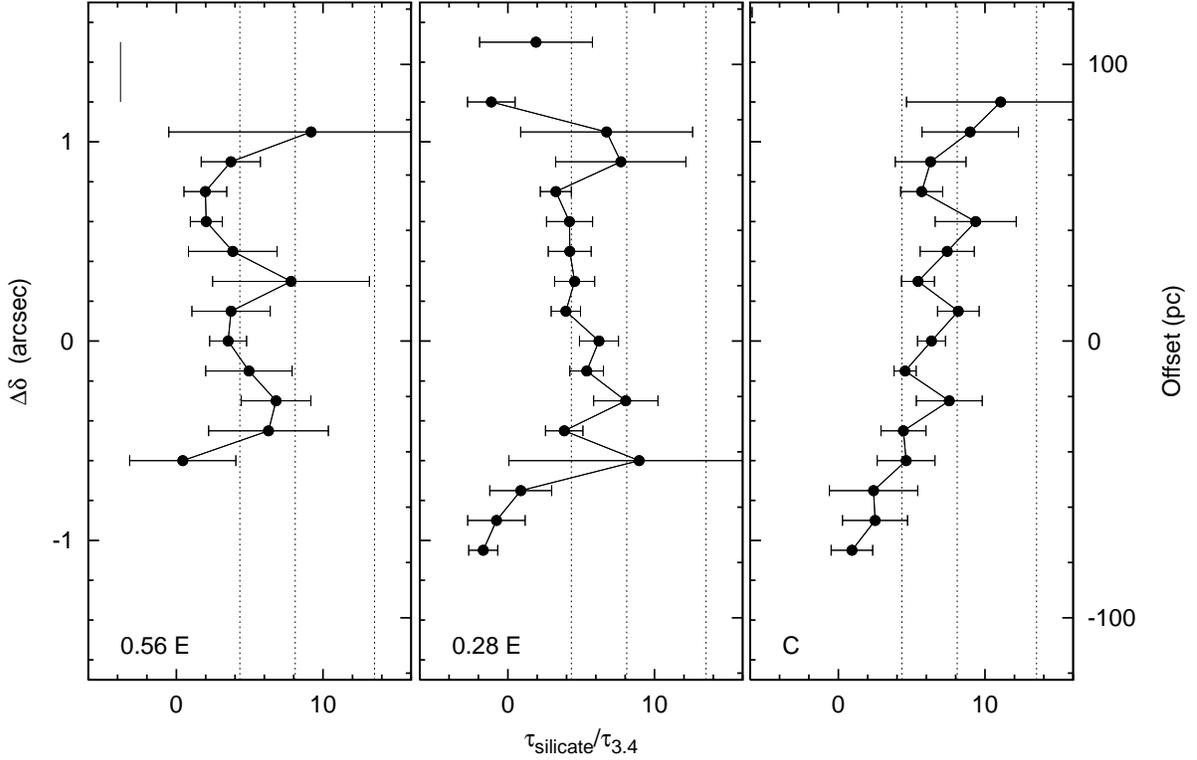}
  \caption[Ratio of optical depths of the dust features]
  {Ratio of optical depths of the dust features.
    The numerator $\tau_{\mbox{\scriptsize silicate}}$ is either
    $\tau_{9.7}$ or $\tau_{10.4}$, whichever had the higher absolute value.
    The vertical dotted lines from left to right show the ratios measured
    on the 3{\farcs}8$\times$3{\farcs}8 aperture
    \citep{Imanishi2000}, on Galactic center sources, and on local
    ISM \citep{Pendleton1994}, respectively.
    Error bars are the 1$\sigma$ uncertainty of each measurement.
    The vertical line at top left shows the spatial resolution (FWHM) of
    the data.
  \label{fig:ratio}}
\end{figure}

Our results show significant spatial variation of the ratio
$\tau_{\mbox{\scriptsize silicate}}/\tau_{3.4}$ (Figure \ref{fig:ratio}),
with a complexity not expected to be produced by a dusty torus.
As shown in Figure \ref{fig:mirspec}, the 9.7~{\micron} silicate feature
is shifted towards longer wavelengths at some locations: optical depth of
the feature is deeper at 10.4~{\micron} than at 9.7~{\micron} ({\S}
\ref{sec:10.4}).
To minimize influence from the wavelength shift, we defined
$\tau_{\mbox{\scriptsize silicate}}$ as either $\tau_{9.7}$ or $\tau_{10.4}$
whichever has the higher absolute value.
At the central peak, the ratio from our observation is between that measured by
\cite{Imanishi2000} and the Galactic values.
In the slit C, the ratio increases from the central peak towards
the north and approaches the Galactic values.
In the eastern slits, the ratio is flatter and closer to that measured
by \cite{Imanishi2000}.
In the south, the ratio is smaller.
Figure \ref{fig:tau} shows that the silicate feature is observed as
emission in the south.

The emission of the silicate feature in the south of the central peak
implies a temperature gradient in the line of sight:
e.g., an optically thin layer of warm dust is present in front of cooler
continuum source.
On the other hand,
in the north, the change of the ratio $\tau_{\mbox{\scriptsize
silicate}}/\tau_{3.4}$ may be explained either with a temperature
gradient in the
line of sight: warmer dust clouds are located behind cooler dust clouds,
or with differences in
dust composition.
Spatially resolved MIR spectra are needed to disentangle the
possibilities.

\subsection{Pie shaped warm dust cloud}
\label{sec:pie}

With the spatial resolution of 26~pc on the sky, the plausible dusty
torus is not detected.
Instead,
we found a pie shaped area
of high MIR emissivity in the north up to about
50~pc from the central engine (Figure \ref{fig:MIRTOS}{\it d}).
The pie shaped area is in the west of the radio jet.
There is no other hint of a channel supplying material to the central
engine in the IR images in the literature and in this study.
Therefore, we speculate that the area might be a channel
that supplies material to the central engine.

The lifetime of this cloud can be estimated assuming that material falls down
into the central engine through the cloud.
The dust mass of the cloud is estimated at around $4.6 \times
10^{5}$~M$_{\sun}$ integrating $\varepsilon_{cont}(\mbox{MIR})$ over the
cloud where $\varepsilon_{cont}(\mbox{MIR}) \ge 10^{-1.6}$ and dividing
with $K_{abs}$ ({\S} \ref{sec:MIRfitting}).
Assuming the dust-to-gas ratio of 100 \citep{Contini2003}, total mass of
the cloud is $\sim 4.6\times 10^7$~M$_{\sun}$.
Mass accretion rate to the central engine is estimated at
0.05~M$_{\sun}$~year$^{-1}$ from total luminosity $L =
2.2\times10^{11}\mbox{~L$_\sun$}$ \citep[at luminosity distance of
15~Mpc]{Telesco1980}
and a typical value of mass-to-luminosity conversion
efficiency of a black hole $L \sim 0.3 c^2 dM/dt$ \citep{Eardley1975}.
The estimated total mass and accretion rate yields a
lifetime of the cloud to be on the order of 10$^9$ years.

\section{CONCLUSION}
\label{sec:Conclusion}

We observed an area of 2{\farcs}8 (200~pc) around the central peak of
NGC 1068 in the MIR
(8.8 -- 12.3~{\micron}) and in the {\it L}-band (3.0 -- 3.9{~\micron}).
The shifted-and-added MIR images have a spatial resolution of
0{\farcs}37 or a projected distance of 26~pc while
the {\it L}-band spectra are taken with a seeing limited spatial
resolution of 0{\farcs}3 or 22~pc.
From these data, we derived grey body parameters: color temperatures of
the continua and emissivities, which
are proportional to column densities of dust emitting the
continua, at each spatial element.
Moreover, optical depths of spectral features of silicate around
9.7~{\micron} and carbonaceous dust around 3.4~{\micron} were also derived
({\S}{\S} \ref{sec:MIRfitting}, \ref{sec:3.4}).

The extended continua over 100~pc to the north and to the south
of the central source are detected both in the MIR and in the {\it L}-band.
We found that
the MIR continuum is mainly emitted from dust in thermal equilibrium with
radiation from the central engine while the {\it L}-band continuum is
emitted by VSGs ({\S} \ref{sec:background_source}).
The observed SED in the MIR suggests that the peak wavelength of the
9.7~{\micron} silicate absorption feature is shifted to a longer
wavelength at some locations ({\S} \ref{sec:10.4}).
The ratio of the optical depths of the silicate and carbonaceous dust
features show complicated spatial distribution ({\S} \ref{sec:dust}).
There is a pie shaped area of enhanced MIR emissivity extending about
50~pc to the north from the central engine.
The morphology of the cloud leads us to speculate that the area is
a channel that feeds material into the central engine
({\S} \ref{sec:pie}).

\acknowledgments
We would like to acknowledge Miwa Goto for help in reducing the IRCS
data.
We would also like to thank the anonymous referee for the helpful
comments which significantly improved this paper.
Part of this work was performed when D.~T. was at Max-Planck-Institut
f\"ur extraterrestrische Physik (MPE).
D.~T. would like to acknowledge the colleagues
at MPE and Takeo Minezaki for helpful comments and inspiring discussions.
Cathy Ishida and Mark Garboden helped us in preparing early versions of
the manuscript.
The data presented in this paper are acquired in the commissioning phase
of the telescope and the instruments.
We are very grateful to all the Subaru staffs who committed themselves
to the telescope and the instruments.

\end{document}